Chromospheric response during the precursor and the main phase of a B6.4 flare on August 20, 2005

Arun Kumar Awasthi,[1] Pawel Rudawy,[2] Robert Falewicz,[2] Arkadiusz Berlicki,[2,3] and Rui Liu[1]

[1]*CAS Key Laboratory of Geospace Environment, Department of Geophysics and Planetary Sciences, University of Science and Technology of China, Hefei 230026, China*[*]
[2]*Astronomical Institute, University of Wroclaw, Kopernika 11, 51-622, Wroclaw, Poland*
[3]*Astronomical Institute, Czech Academy of Sciences, 25165 Ondrejov, Czech Republic*

ABSTRACT

Solar flare precursors depict constrained rate of energy release contrasting the imminent rapid energy release which calls for different regime of plasma processes to be at play. Due to subtle emission during the precursor phase, its diagnostics remain delusive, revealing either the non-thermal electrons (NTEs) or the thermal conduction to be the driver. In this regard, we investigate the chromospheric response during various phases of a B6.4 flare on August 20, 2005. Spatio-temporal investigation of flare ribbon enhancement during the precursor phase, carried out using spectra-images recorded in several wavelength positions on the H$\alpha$ line profile, revealed its delayed response (180 seconds) compared to the X-ray emission, as well as sequential increment in the width of the line-profile which are indicative of a slow heating process. However, energy contained in the H$\alpha$ emission during the precursor phase reach as high as 80% of that estimated during the main phase. Additionally, the plasma hydrodynamics during the precursor phase, as resulted from the application of a single-loop one-dimensional model, revealed the presence of power-law extension in the model generated X-ray spectra, with flux lower than the *RHESSI* background. Therefore, our multi-wavelength diagnostics and hydrodynamical modeling of the precursor emission indicates the role of a *two-stage* process. Firstly, reconnection triggered NTEs, although too small in flux to overcome the observational constraints, thermalize in the upper chromosphere. This leads to the generation of a slow conduction front which causes plasma heating during the precursor phase.

1. INTRODUCTION

Solar flares are violent phenomena occurring in solar atmosphere and release $\sim 10^{32}$ ergs energy in typically $10^3$ seconds. In principle, such phenomena occur in active regions having complex magnetic field configuration (Choudhary et al. 2013; Jain et al. 2011). Moreover, emission from a strong flare can be recorded from gamma-rays to radio wavebands i.e. covering almost entire electromagnetic spectrum (Fletcher et al. 2011). Thorough analysis of such a complex and violent phenomenon needs extended theoretical and numerical modelling based on state-of-the-art resolution observations and codes. Despite enormous progress in solar flare investigations, the most-applicable 2D model of the energy release in solar flares till date is still CSHKP model proposed by Carmichael (1964), Sturrock (1966), Hirayama (1974), and Kopp & Pneuman (1976). Modifications inspired by observational studies along with the theoretical simulations over the years resulted in a standard (unified) model of the flares (Dennis & Schwartz 1989; Shibata 1999).

Corresponding author: Arun Kumar Awasthi
arun.awasthi.87@gmail.com

[*] Part of the work has been done while employed at 'Astronomical Institute, University of Wroclaw, Kopernika 11, 51-622, Wroclaw, Poland'

According to the standard model of solar flares, magnetic reconnection is understood to be the mechanism working as the engine for converting the magnetic energy into the kinetic energy of the charged particles (mostly electrons). Electrons accelerated in this manner reach the chromosphere guided by magnetic field of the flaring region and dump its energy in the dense chromosphere while a tiny part (roughly $10^{-5}E_{tot}$) of the energy is radiated in X-ray wavelength in the framework of thick-target bremsstrahlung (Brown 1972). Most of the energy of the charged particle is also utilized in heating the chromosphere, the transition region, and the coronal plasma attached to the reconnected field lines, as well as sometimes even the photosphere. Subsequently, a sudden increase in the energy of small volume/kernel of chromosphere results in chromospheric evaporation/ablation. The ambient charged particles thus start filling the loop guided by the flare loop magnetic field and undergo thin-target bremsstrahlung resulting in low-energy (soft) X-ray (SXR) emission. Integrating the temporal evolution of energy release as explained above, SXR emission should always be accompanied by high-energy (hard) X-ray (HXR) counterpart. On the contrary, precursor SXR enhancement has been recorded well before the onset of impulsive HXR emission as often as in 90% of the flares studied by Veronig et al. (2002). Therefore, temporal evolution of a typical flare comprehending the aforementioned physical process may be divided in three sub-intervals namely the precursor phase, the impulsive phase and, the gradual (decay) phase. The precursor phase of the flare is most commonly referred to gradual small-scale enhancements, prominently seen in the SXR and EUV wavebands, prior to the onset of rapid and intensive emission recorded in the HXR and radio wavebands, termed as the impulsive phase. On the other hand, the post-impulsive phase activity is collectively termed as the gradual phase of the flare. While multi-wavelength diagnostics of precursor activities reveal vast variety of process including 1) discrete and localised X-ray brightenings with the possibility to be detected as early as 50 min before the impulsive phase of the flare and filament acceleration (Ohyama & Shibata 1997; Chifor et al. 2007), 2) slow-rise in the filament (Martin 1980; Chifor et al. 2006; Joshi et al. 2011) as well as partial filament eruption (Awasthi et al. 2014), 3) reconnection within various branches of a multi-flux-rope system during the precursor phase of a confined flare (Awasthi et al. 2018), and 4) oscillation of a magnetic flux-rope (MFR) during the precursor phase suggestive of magnetic restructuring during this phase (Zhou et al. 2016), underlying mechanism which gives rise to these small intensity-scale activities as well as its role in the impulsive phase of the flare is still debated.

Several studies have been carried out in order to understand the origin of precursor phase emission. Battaglia et al. (2009) suggested the thermal conduction driven chromospheric emission to be a possible mechanism of the precursor SXR enhancement. Awasthi et al. (2014), in their study of a solar flare event well-observed in multi-wavelength band, presented observational signature of conduction front responsible for precursor phase emission. However, Falewicz et al. (2011) and Falewicz (2014) found the energy contained in the reconnection driven non-thermal electrons (NTEs) to be sufficient to produce SXR emission during the precursor phase. They argued the insufficient instrumental sensitivity to be responsible for the absence of HXR emission during the precursor phase. Altyntsev et al. (2012) analyzed microwave emission during the precursor as well as impulsive phase and suggested that although non-thermal emission remained present during the precursor phase, it was not sufficient to explain the observed SXR enhancement. On the other hand, regardless of the higher sensitivity of the radio instruments in comparison to the existing X-ray detectors, Benz et al. (2017) reported a radio-quiet preflare phase with three well-developed SXR extrusion as recorded by *RHESSI*. Battaglia et al. (2014) attempted to search for chromospheric response due to the conduction front during the precursor phase of a flare using high spatial and temporal cadence observations for one well-observed flare. Their study resulted in the absence of noticeable signatures of the energy transported through conduction at the chromospheric heights. Therefore, it is evident that the study of energy release processes during the precursor phase is yet to reach a consensus.

In-depth investigation of chromospheric response is crucial to shed light on the physical process taking place at different layers of the solar atmosphere leading to the precursor excursions and their association with the imminent impulsive energy release during the main phase of the flare. Complex dynamical activities of the plasma giving rise to the flare emission, particularly in the chromosphere, can be probed through deriving the characteristics of H$\alpha$ line profile (Zarro et al. 1988; Graeter & Kucera 1992; Heinzel et al. 1994; Druett et al. 2017) in the form of shift in the line-center, and the line-width. Canfield et al. (1984) investigated the effect of non-thermal electron beam and coronal pressure enhancement on the synthesized H$\alpha$ line profiles and suggested that increased coronal pressure can cause the line profile broadening, and increase in the total intensity of the H$\alpha$ line. On the other hand, only a large flux of high-energy non-thermal electrons may result in the H$\alpha$ profiles having non-Gaussian broad wings. Moreover, their study could associate the line central reversal effects with the heating rate by non-thermal electrons. On the contrary, thermal conduction mechanism may result in a reduced H$\alpha$ line width and total intensity of the profiles.

Huang et al. (2014) investigated the trigger for observed sequence of pre-eruptive events leading to a filament eruption as well as coronal mass ejection. Their study revealed draining plasma into the loop foot-point to be associated with the precursor phase and the subsequent filament eruption. Moreover, while several investigations have shown the blue asymmetry during the early rise phase of the flare (see Heinzel et al. (1994) and the references therein), the underlying physical processes giving rise to the same remains illusive. In particular, although Heinzel et al. (1994) proposed the electron beam heating with the return current to be causing the early stage blue-shift in the H$\alpha$ line profile, similar effect in the line profile have been reported in the investigation of Graeter & Kucera (1992), who found the erupting filament to be the cause. Further, Kuridze et al. (2015) emphasized the role of steep gradient in the velocity of emitting plasma in shifting the wavelength of maximum opacity, thus adding another dimension of complexity in the implications obtained merely by the shift in the H$\alpha$ line profile. Therefore, although the quantification of H$\alpha$ line profile provides comprehensive information of the dynamical activities during various stages of the flare evolution, a careful investigation employing co-temporal spectra-images such as those recorded with the Multi-channel Subtractive Double Pass (MSDP) imaging spectrograph (Mein 1991; Rompolt 1990) is inevitable as performed in the present paper.

In this paper, we present a multi-wavelength diagnostics of the chromospheric response during various phases of a small B6.4 flare of August 20, 2005. In section 2, a brief discussion on the observations applied in the present study and the respective instrument is presented. Section 3 deals with the multi-wavelength data analysis while section 4 presents an in-depth analysis of the spectra-images produced by MSDP-type spectrograph covering H$\alpha$ line center and wings. Section 5 provides a comparative overview of the thermal and non-thermal energy content estimated from the X-ray and H$\alpha$ observations during various phases of the flare. In section 6, we present the results of single-loop one-dimensional hydrodynamic model in context of the observations during various phases of the flare. Section 7 offers the conclusions and insights gained from the present study.

## 2. OBSERVATIONS AND INSTRUMENTATION

We study the spatial, temporal and spectral evolution of chromospheric response during the precursor and main phases of SOL2005-08-20T08:09 event, a B6.4 intensity class flare, which occurred in AR10798 (S10W33) on August 20, 2005. In this regard, we analyze multi-wavelength observations provided by the following instruments.

### 2.1. MSDP

Spectra-images of the investigated flare (i.e. two-dimensional images convolved with the H$\alpha$ line spectra; see Mein (1977) for details) have been collected with the Large Coronagraph (Gnevyshev et al. 1967; Rompolt & Rudawy 1985) equipped with the Multi-channel Subtractive Double Pass (MSDP) imaging spectrograph (Mein 1991; Rompolt 1990) at the Bialkow Observatory of the University of Wroclaw, Poland. The Large Coronagraph (LC) is a classical Lyot-type instrument with a 51 cm diameter main objective and nearly 14.5 m effective focal length which can also be used as a chromosphere-graph (without the artificial Moon). The rectangular entrance window of the MSDP spectrograph is located at the coude-focus of the LC which covers an equivalent area of 325 $\times$ 41 arcsec$^2$ on the sky plane delimiting the effective field of view (FOV) of the whole system. Because of the limited FOV, scans of various lengths of 15-18 spectra-images are usually obtained to cover the whole flare region as well as the surrounding area. In this way, effectively an area of $\sim$ 300 $\times$ 360 arcsecs$^2$ in the plane of sky is covered. The spectrograph has a nine-channel prism-box (Mein 1991; Rompolt 1990) creating d$\lambda$ = 0.4 Å steps in wavelength between consecutive nine "channels" at the spectra-images. All nine channels at the individual quasi-monochromatic spectra-image (waveband = 0.06 Å) show exactly the same area on the Sun, but each of these is recorded in a slightly shifted wavelength band in relation to others in the vicinity of the H$\alpha$ line (the width of the wavelength band of a single channel is $\Delta\lambda$ = 0.32 nm). The spectra-images have been recorded with a 12-bit Photometrics KAF1400 CCD camera using 0.57 arcsec per CCD pixel sampling and 40 ms exposure time. The time interval between two consecutive exposures has been nearly 2.8 s, the effective time step between consecutive scans varied between 55 s and 68 s, depending upon the change in observing parameters. The DCF77 time signal receiver was used to record the precise time of each exposure. For each scan we constructed a set of thirteen narrow-band nearly monochromatic images of the whole region (up to $\pm$ 1.2 Å from the H$\alpha$ line center), as well as the H$\alpha$ line profile at each pixel in the field of view. The spatial resolution of the obtained images is limited by seeing, on average to about 1 arcsec. We have used Meudon spectroheliograph full-disk observations obtained at 06:20:16 UT as a reference to derive actual Solar X-Y coordinates of high-resolution MSDP observations. Additionally, *SOHO*/MDI full-disk photospheric continuum observation, recorded at 06:24:00 UT, is used for cross-checking the alignment. In the present investigation, we analyze MSDP observations recorded during 06:36 UT - 09:00 UT, covering various phases of the flare. However, bad weather conditions during 07:10 UT - 07:25 UT led to a gap in the observations for this duration.

### 2.2. *RHESSI*

X-ray emission recorded by the *Reuven Ramaty High Energy Solar Spectroscopic Imager* (*RHESSI*; Lin et al. (2002)) during the flare is analyzed to deduce the spatial, temporal and spectral evolution of X-ray source. *RHESSI* records X-ray emission in 3 keV - 20 MeV energy band employing nine Germanium detectors with a temporal cadence of 4 second, and the energy resolution ranging between 1 to 5 keV. In case of the analysed flare *RHESSI* recorded the X-ray counts in the 'A1' attenuator state over the entire flare. Due to *RHESSI* night, a data gap occurred between 06:45 - 07:29 UT. We employ data recorded with detectors 2F, 3F, 4F, 5F, 6F, 8F, and 9F for the synthesis of spectra and images. Figure 1 shows the temporal evolution of X-ray emission of the flare in various energy bands covering 6-25 keV.

### 2.3. *SOHO/EIT*

Extreme ultra-violet observations in 195Å band, recorded during the flare by Extreme Ultraviolet Imager Telescope (EIT; (Delaboudinière et al. 1995) on-board the *Solar and Heliospheric Observatory* (*SOHO*) have been analyzed. Temporal cadence of such observations is 12 minute while the spatial resolution is $\sim$ 2.6 arc second. Temperature coverage of observed emission from this instrument is in the range of 0.08-2 MK. Temporal evolution of background-subtracted EUV emission during the flare is plotted in Figure 1.

## 3. MULTI-WAVELENGTH DIAGNOSTICS OF THE FLARING PLASMA DURING THE PRECURSOR AND THE MAIN PHASE

Multi-wavelength emission recorded by various space- and ground-based instruments has been investigated. As following, we present the temporal, spatial and spectral diagnostics of the emission corresponding to various phases of the flare.

### 3.1. *Spectral and spatial evolution of the X-ray emission*

We analyse the spatial, temporal and spectral evolution of X-ray sources as seen by *RHESSI* during the flare. Intensity profiles of X-ray emission in various energy bands covering 6-25 keV are plotted in Figure 1. Although enhancement in the energies >12 keV appeared unambiguously only after $\sim$ 08:04 UT in *RHESSI* observations, successive and rapid enhancement in the 0.5-4 Å emission (higher energy channel of *GOES*) (Figure 1b) has been recorded since 07:45 UT which marks the onset of the main phase of the flare. On the other hand, low-energy X-ray emission in the form of several episodes of small-scale humps is recorded in *RHESSI* as well as in 1-8Å wavelength bands as early as since $\sim$ 06:00 UT. A notable trend in the 1-8 Å intensity profile since 06:00 UT is a slow but steady increase in the X-ray emission level despite several intermittent emission peaks, which led us to consider 06:00-07:45 UT to be the precursor phase of the flare investigated in this paper.

Thermal and non-thermal characteristics of the flare plasma are diagnosed by investigating the X-ray emission recorded by *RHESSI* mission. For this, firstly we prepare X-ray spectra with a non-uniform temporal and energy binning in order to achieve sufficiently enough count statistics during various phases of the flare. During 06:00-06:45 UT, and 07:46-07:56 UT, covering the precursor, and a part of the rise phase durations of the flare, the spectra have been prepared with the signal averaged over a period of 120 seconds, while the time duration for averaging the records is kept to 32 seconds for the same corresponding to the main phase during 07:56-08:27 UT. Further, the spectra is prepared with the energy binning of 0.3 keV in the 6-10 keV energy band and the same has been increased to 1 keV for the emission in the energies >10 keV. As the X-ray emission originating from the coronal plasma is an aggregated response of thermal and non-thermal processes, we perform a forward-fit of the background-subtracted *RHESSI* observations with a theoretical spectrum prepared by combining the iso-thermal (`v_th.pro`) and thick-target bremsstrahlung models (`v_thick2.pro`) available within the SPectral EXecutive (SPEX) package of SolarSoft[1](Awasthi et al. 2016). Line and continuum emissions in the iso-thermal model are generated using the CHIANTI atomic database (Dere et al. 1997; Landi et al. 2013) while the non-thermal electrons in the spectral shape of power-law function (Brown 1971) are implemented in the thick-target model. The observed X-ray spectra serves as an input to the fit-procedure in SPEX which is performed through the `mcurvefit.pro` routine, based on the nonlinear least-squares Levenberg-Marquardt fitting algorithm. In this, by varying the inherent parameters of the models in an iterative manner, a model spectrum is derived which best-fits the observations. The goodness of the fit is determined by estimating the reduced $\chi^2$ value with the aim to minimize it to 'unity'. The spectral fit enabled us to estimate the flare plasma parameters characterizing thermal emission namely temperature (T) and emission measure (EM), while those representing non-thermal emission are electron flux, negative spectral index ($\delta$) and lower energy cutoff. Temporal evolution of the aforementioned parameters during various phases of the flare is shown in figure 2. Moreover, the uncertainties in the fit parameters have also been estimated within SPEX package by making use of curvature matrix of the $\chi^2$ hypersurface in the parameter space assuming the parameters to be well-defined and have symmetric uncertainties of local gaussian shape (Ireland et al. 2013; Warmuth & Mann 2016a) as shown in the form of error bars over the respective quantities in the Figure 2.

---

[1] http://www.lmsal.com/solarsoft/

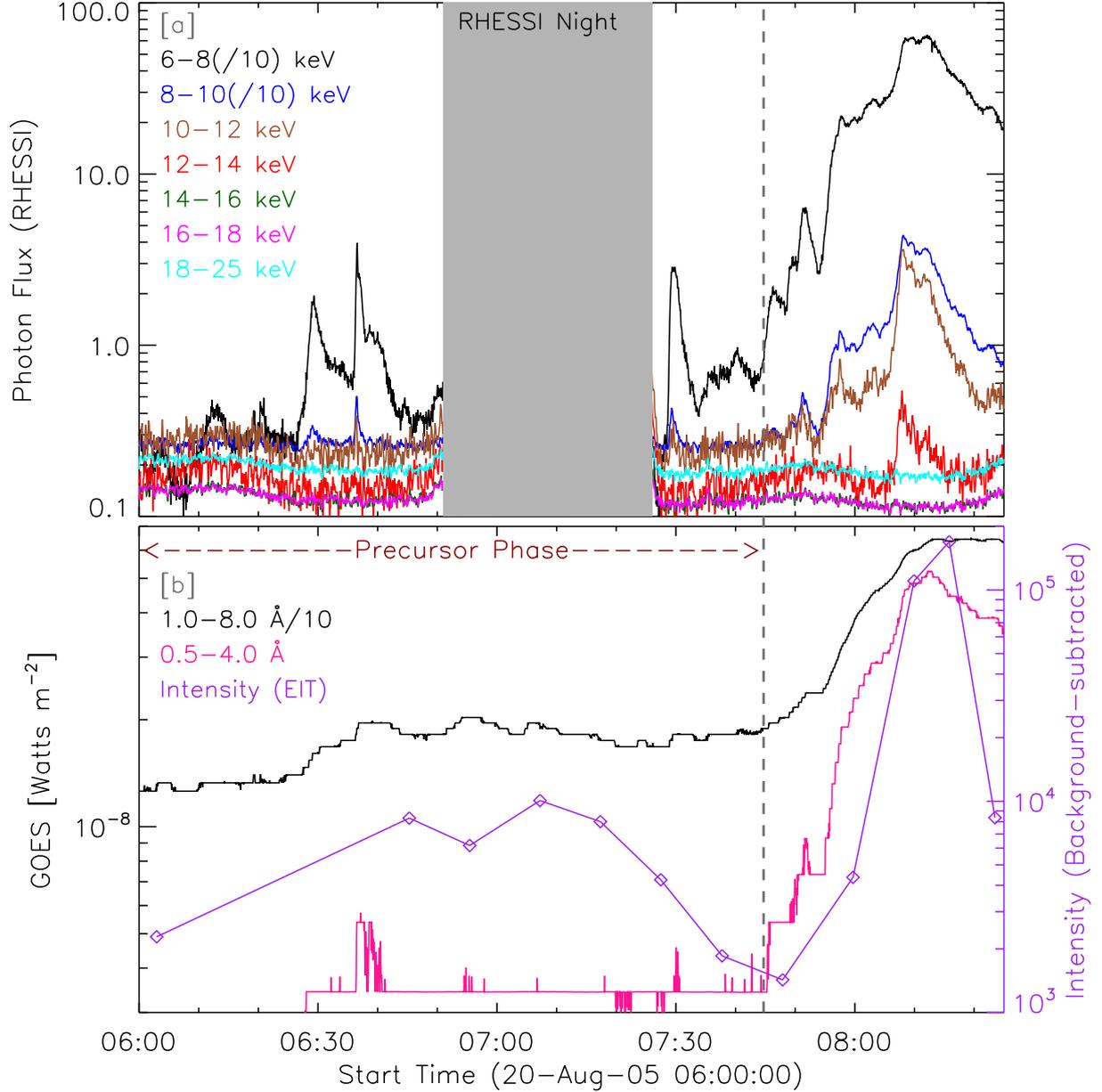

**Figure 1.** Evolution of X-ray emission indicating various small scale enhancements characterized as precursor activities before the onset of the main phase of the flare, commencing at 07:45 UT. [a]. X-ray photon flux in various energy bands within 6 and 25 keV as recorded by *RHESSI*. Light grey shaded area denotes the data gap due to *RHESSI* night as well as its passage through the SAA region. [b]. *GOES* observations of the X-ray emission in 0.5-4 Å (pink) and 1-8 Å (black) along with the background-subtracted 195 Å emission (purple), estimated from the images taken by *SOHO*/EIT instrument.

During the precursor phase, plasma temperature varies in the range of 9-14 MK while during the main phase it ranges between 12-24 MK. Emission measure (EM) of the plasma reaches to a value of $5 \times 10^{47}$ cm$^{-3}$ during the precursor phase, which is retained throughout the main phase. Due to poor count rate recorded during the precursor phase, the temperature and EM estimations suffer higher uncertainties than that during the main phase. Next, non-thermal electron beam parameters, namely total electron flux, power-law index and low-energy cut-off values, as plotted in figure 2, vary in the range of 1-7 $\times$ 10$^{35}$, 7-14, and 8-10 keV, respectively. We also plot the temporal evolution of observed X-ray emission in 6-10, 10-14 and 14-25 keV energy bands in the bottom panel of Figure 2 for reference.

Next, we investigate the evolution of X-ray source morphology during various phases of the flare using the X-ray images in 6-10 keV, 10-12 keV, 12-15 keV, and 15-20 keV energy bands, reconstructed with the CLEAN algorithm (Hurford et al. 2002). Figure 3 shows the evolution of X-ray sources along with the magnetic-field topology and the loop morphology. The magnetograms obtained from Michelson Doppler Imager (MDI; Scherrer et al. (1995)) instrument on-board *SOHO* mission have been applied. As the *SOHO*/MDI magnetograms with a high time cadence of one minute were available only until 08:00:00 UT, evolution of magnetic-field parameters has only been comprehensively studied during the precursor phase of the flare. This investigation revealed a persistently increasing trend in both the positive as well as negative flux values until 06:18 UT, associated well in time with the onset of X-ray precursor enhancement recorded in *GOES* observations. Subsequently, the flux cancellation is evident, although much clear in the positive flux evolution, plotted in the Appendix Figure 1 (Appendix A). It might suggest magnetic reconnection to be responsible for triggering the energy release during the precursor phase, in agreement to that revealed in the investigation of Wang et al. (2017) of two precursor excursions of a M6.5 flare on 22-June-2015. EUV images obtained in 195 Å during the flare have been analyzed in order to distinguish the morphological changes in the loop connectivity from the precursor to the main phase of the flare. Temporal evolution of background-subtracted emission in 195 Å is plotted in Figure 1 which reveals enhanced EUV emission during the precursor phase. The coronal loops connectivity in conjunction with the photospheric magnetic field topology is investigated through the iso-intensity contours of magnetic-field value 800 Gauss of both the positive and negative polarities plotted over the 195 Å EUV images as shown in Figure 3. Co-spatial emission during the precursor and main phase is seen from the time sequence of EUV images. We further over-plot the contours of 80% of the maximum intensity of X-ray images in 6-10, 10-15 and 15-35 keV. From this multi-wavelength emission representation, the flare activity in the corona appears to be predominately concentrated over the western (positive) polarity of the active region. Further, systematic shift in the centroid location of X-ray emission towards the positive polarity is found from the precursor to main phase.

## 4. CHROMOSPHERIC RESPONSE DURING VARIOUS PHASES OF THE FLARE

The quasi-monochromatic images reconstructed using MSDP data in the H$\alpha$ $\pm$ 1.2Å wavelength band have been analyzed to study the chromospheric response during various phases of the flare. Figure 4 shows the time sequence of images obtained in H$\alpha$ line centre as well as in $\pm$ 0.6 Å in the line wings at the instances 06:36:41 UT (precursor), 08:09:17 UT (flare maximum) and 08:20:23 UT (gradual phase). The iso-contours of 80% of the maximum intensity of the co-temporal X-ray images in 6-10, 10-15, and 15-35 keV energy bands are also drawn on the images. In addition, for reference purpose, we have also drawn iso-intensity contour of magnetic-field strength 800 Gauss as shown in the top panel of the Figure 4.

Figure 4 shows that during the precursor phase (panel [a]), the emitting regions in the H$\alpha$ line centre and in the soft X-ray emission (6-10 keV) are spatially linked. Predominant emissions in H$\alpha$ waveband as well as in the SXR energy band are found to be originating close to the positive polarity. Further, at the maximum of the main phase, i.e. at 08:09:17 UT (panel [b]), we note enhanced H$\alpha$ brightenings in the form of well-defined kernels. A systematic shift of the source centroid in the H$\alpha$ as well as in X-ray wavelengths towards positive polarity while transition of the flare from the precursor (panel [a]) phase to the gradual (panel [c]) phase is found. Comparative morphological analysis of the H$\alpha$ emission during the precursor phase revealed diffused spatial distribution of the sources while those appearing in the form of distinct kernels during the impulsive phase of the flare. Further, the dynamics of the excursions tend to sequentially shift westwards towards the positive polarity similar to that exhibited by the X-ray contours. Irrespective of several minor activities in the filament evolution associated with the flaring region, any sequential filament activity associated with various phases of the flare remained ambiguous, nevertheless, this implication is limited by the moderate spatial-resolution of MSDP observations. A quantitative investigation of the time evolution of Hα line profile in various phases of the flare follows.

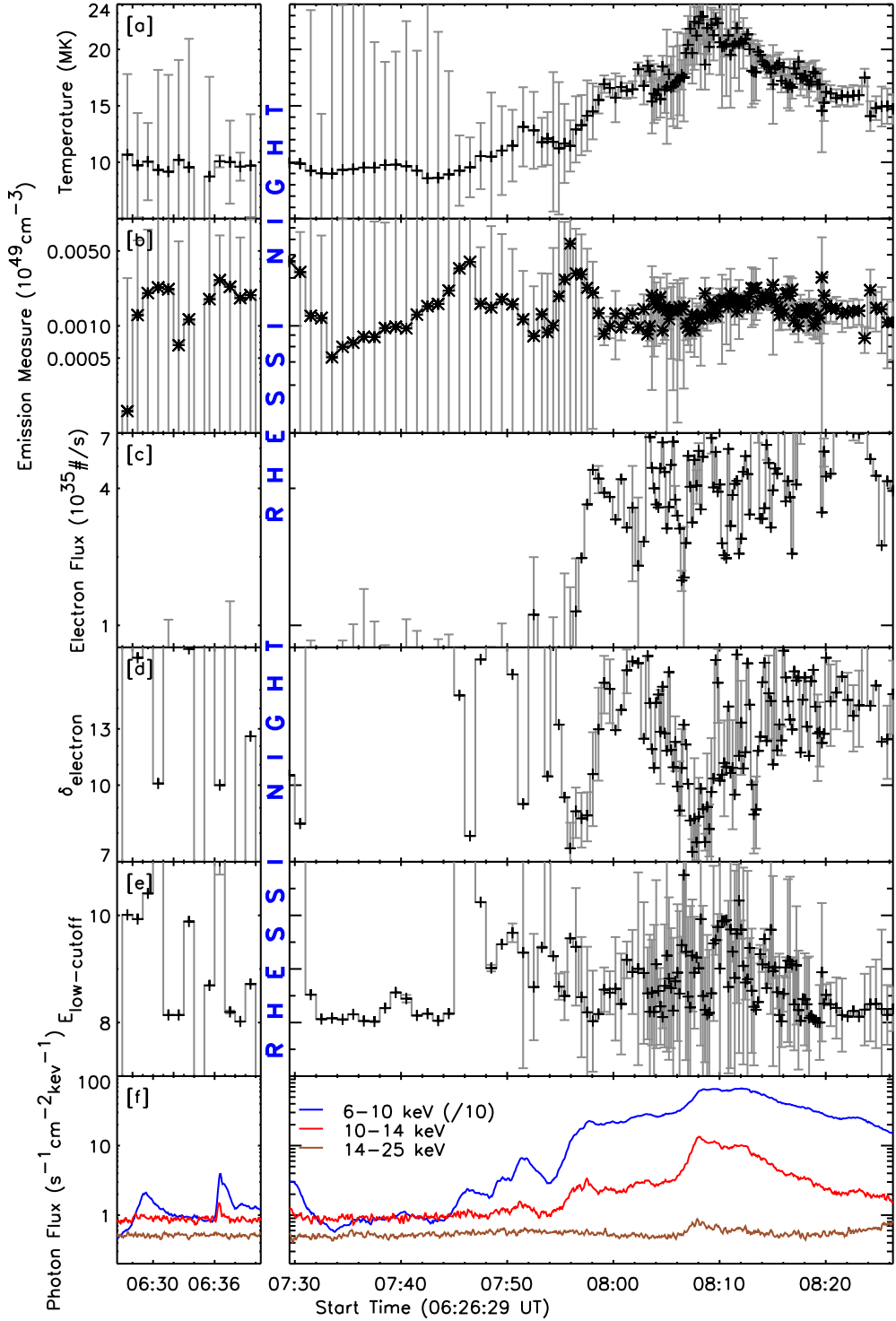

**Figure 2.** Evolution of plasma parameters namely temperature, emission measure, electron flux, power-law index ($\delta_{electron}$), and cut-off energy (panel [a] - [e]). Panel [f] shows the temporal evolution of observed X-ray emissions in 6-10, 10-14 and 14-25 keV energy bands. Photon flux in 6-10 keV is reduced by one order for better visibility of the intensity evolution in high energy bands.

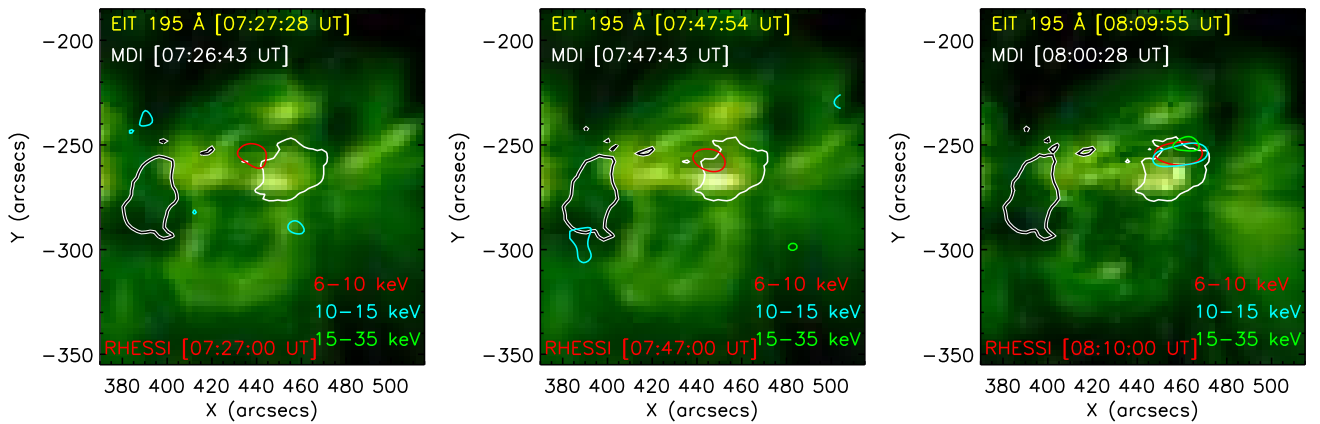

**Figure 3.** Evolution of coronal activities during various phases of the flare along with the photospheric magnetic field morphology. Sequence of *SOHO*/EIT 195 Å images at various stages of the flare are shown. Co-temporal contours of 80% of the maximum intensities from X-ray images in 6-10, 10-15 and 15-35 keV are drawn in red, cyan and green, respectively. Positive (white) and negative (black) polarities are also drawn in the form of iso-intensity contours of 800 Gauss strength.

### 4.1. *Spatio-temporal diagnostics of the Hα line profile in various phases of the flare*

In order to quantify the emission in the Hα line centre as well as in the wings, we investigate the spectral evolution of flare ribbon in several individual sub-regions denoting the areas covering significant activity during various phases of the flare. As shown in the panel [a] of the Figure 5, we have chosen 15 regions of interest (ROIs) of circular shape. The line profile of the quiet Sun ($I_{quiet}[\lambda, t]$) corresponding to each instance of the observation is derived by averaging the intensity within the two rectangular regions (panel [a]; white box), away from the flaring region, from the sequence of images recorded at the respective time (t) in different wavelength positions across the Hα line. Next, we derive intensities corresponding to various wavelength positions of the Hα line profile by averaging a matrix of dimension 3 × 3 pixels centered at the position of maximum intensity within the ROI. In this way, we have derived a time evolution of Hα line profile corresponding to all the ROIs for the entire duration of the flare. Hα intensities have been standardised to physical units (erg s$^{-1}$ sr$^{-1}$ cm$^{-2}$ Hz$^{-1}$) using the quiet-sun reference spectrum of David (1961) (see Appendix B). The investigation of the intensity evolution for all the 15 sub-regions corresponding to various locations over the flare ribbon can be categorized in two distinctive trends where the first exhibits predominant activity during the precursor phase, while the latter shows enhancement during both the precursor as well as main phases of the flare as shown in the appendix figure 2 and discussed in the appendix C. For further detailed investigation, we concentrate on the activities within ROIs 9, 11 and 13 which comprehend the evolution trends derived from all the 15 ROIs. The time profile of the intensity in the Hα line center for the aforementioned ROIs is plotted in the panels [b], [c] and [d] of the Figure 5, respectively.

Time profile of the Hα line center intensity corresponding to ROI 9 has evolution similar to SXR light-curve during the precursor phase of the flare (Figure 5b). On the other hand, intensity evolution in the Hα line center corresponding to ROI 11 (Figure 5c) is found to be well-correlated with the precursor as well as the main phase of the flare. The intensity profile corresponding to the ROI 13 (Figure 5d) exhibits rapid growth during the maximum of the impulsive phase of the flare, while that during the precursor and the rise phase of the flare is relatively less in comparison to that estimated from ROI 9 and 11. Therefore, as discussed above, we consider the ROIs 9, 11 and 13, as representative cases for different phases of the flare for further investigation.

Time sequence of Hα line profile evolution corresponding to ROIs 9, 11 and 13, presented in the Figure 6, enables us to make a comparative investigation of the chromospheric response during various phases of the flare. In view of

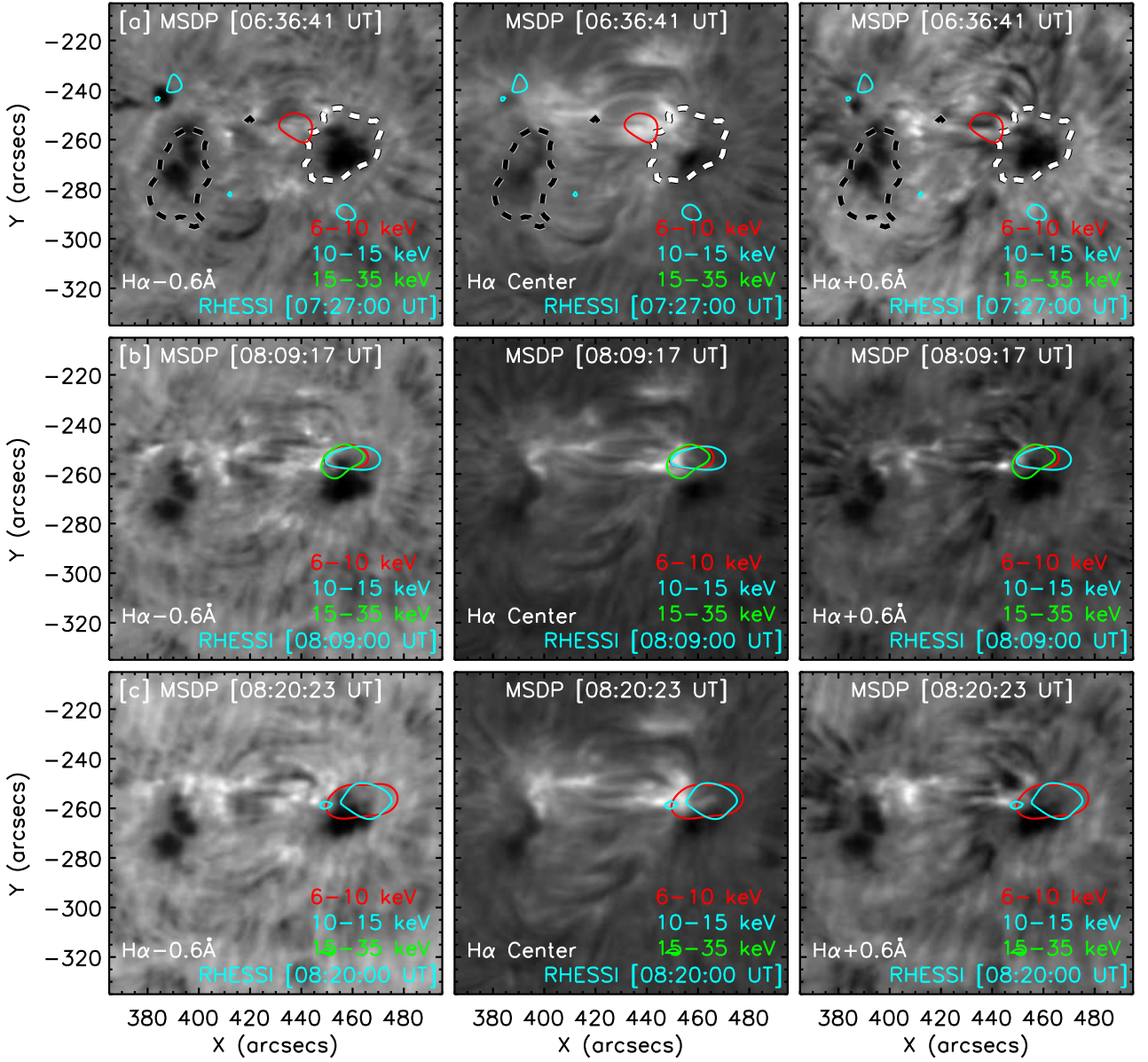

**Figure 4.** Chromospheric response during various phases of the flare as seen in the MSDP spectra-images. Sequence of images in the Hα line center, and in Hα± 0.6 Å wings taken at 06:36:41 UT, 08:09:17 UT and 08:20:23 UT are shown in top, middle and bottom panels, respectively. Iso-contours of 80% of the maximum intensities of X-ray images in 6-10 (red), 10-15 (cyan) and 15-25 (green) keV are drawn. Magnetic-field topology in the form of iso-intensity contour of 800 Gauss magnetic-field strength corresponding to both the polarities (positive (white); negative (black)) is over-plotted in the panel [a].

quantifying the profiles in terms of line broadening, we first calculate the net emission by subtracting the observed profile with the co-temporal quiet-sun profile (Falewicz et al. 2017). Resultant net emission profile is then fitted with the Gaussian function employing the 'gaussfit' procedure available in IDL. The net emission (shown as dotted symbols) as well as best-fit Gaussian function (full-line) for several time instances are plotted in the panels [d]-[f] of figure 6. The full-width-half-maximum (FWHM) of the best-fit to Hα emission profile is considered as the line-width. From the observations corresponding to ROI 9, we find the widths of the line profiles to be sequentially increasing from 0.87

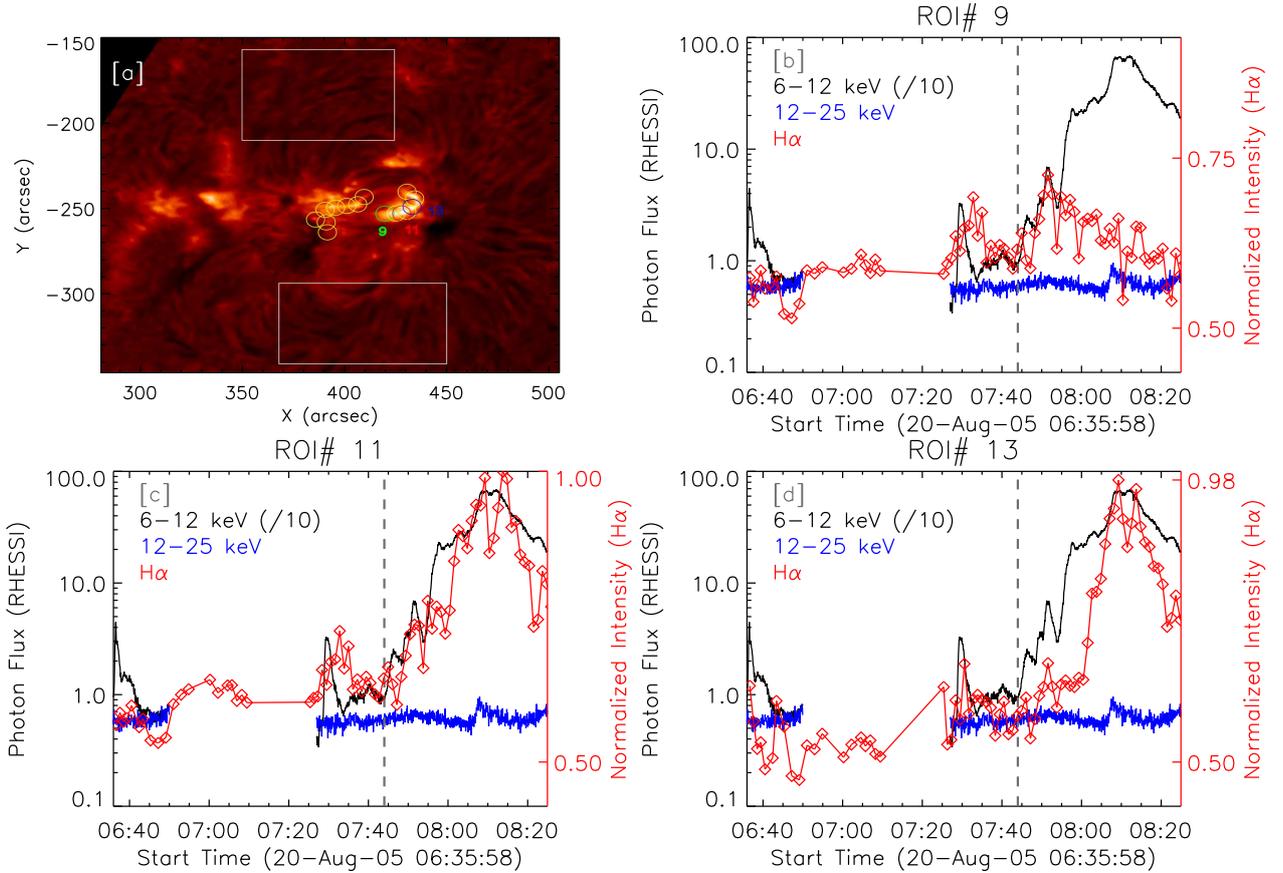

**Figure 5.** Spatio-temporal analysis of the flare ribbon through the spectra-images obtained by MSDP. Fifteen circular shaped sub-regions of interest (ROIs) over the entire flare ribbon, considered for the investigation of the Hα line profile are drawn in yellow in panel [a]. Regions have been sequentially numbered from the left to the right side while only ROIs 9 (green), 11 (red) and 13 (blue) have annotated as those are considered further for detailed analysis. The two rectangles, covering the quiet-sun, at the top and bottom of the flaring region, denote the area over which the quiet-sun profile is estimated. Panels [b], [c] and [d] show the temporal evolution of emission recorded in the Hα line centre for ROIs 9, 11 and 13, respectively. Time profiles of the X-ray photon flux in 6-12 and 12-25 keV are also plotted. The dotted grey line represents the onset of the main phase of the flare.

Å to 1.08 Å from the precursor phase to the rise phase of the flare. On the other hand, such trend in the line-width (broadening) is not evident in the profiles obtained from ROIs 11 and 13. One of the possible interpretations of the line profile broadening phenomena is the enhanced pressure in the flare loop (Canfield et al. 1984) hence further investigation of the MSDP spectra-images is crucial in this context as performed following. During the maximum of the impulsive phase of the flare, the Hα line profile corresponding to the ROIs 11 and 13 show red-shifted emission profiles. Since the location of ROI 11 and 13 is co-spatial to the HXR emission centroid (see Figure 4b) during the maximum of the impulsive phase, we argue such profile to be caused by the intense heating of the chromosphere by the NTEs (Canfield et al. 1990). On the contrary, the line profiles for ROI 9 have never turned in a complete emission profile during the whole period of investigation.

We estimate the relative excess emission in the Hα line centre as well as in the wings using the following equation:

$$I_{excess}(\lambda, t, ROI) = \frac{I_{obs}(\lambda, t, ROI) - I_{qs}(\lambda, t)}{I_{qs}(\lambda, t)} \quad (1)$$

where $I_{obs}(\lambda, t, ROI)$ is the intensity recorded in particular wavelength λ at time 't' within the ROI whereas $I_{qs}(\lambda, t)$ is the intensity of the quiet-sun corresponding to the wavelength at the instance same as that of $I_{obs}$. The temporal evolution of the excess in intensity for the Hα line center as well as ± 0.6 Å for the ROIs 9 and 11 is shown in figure 7. The excess intensity profile for ROI 13 exhibits the evolution similar to that of ROI 11.

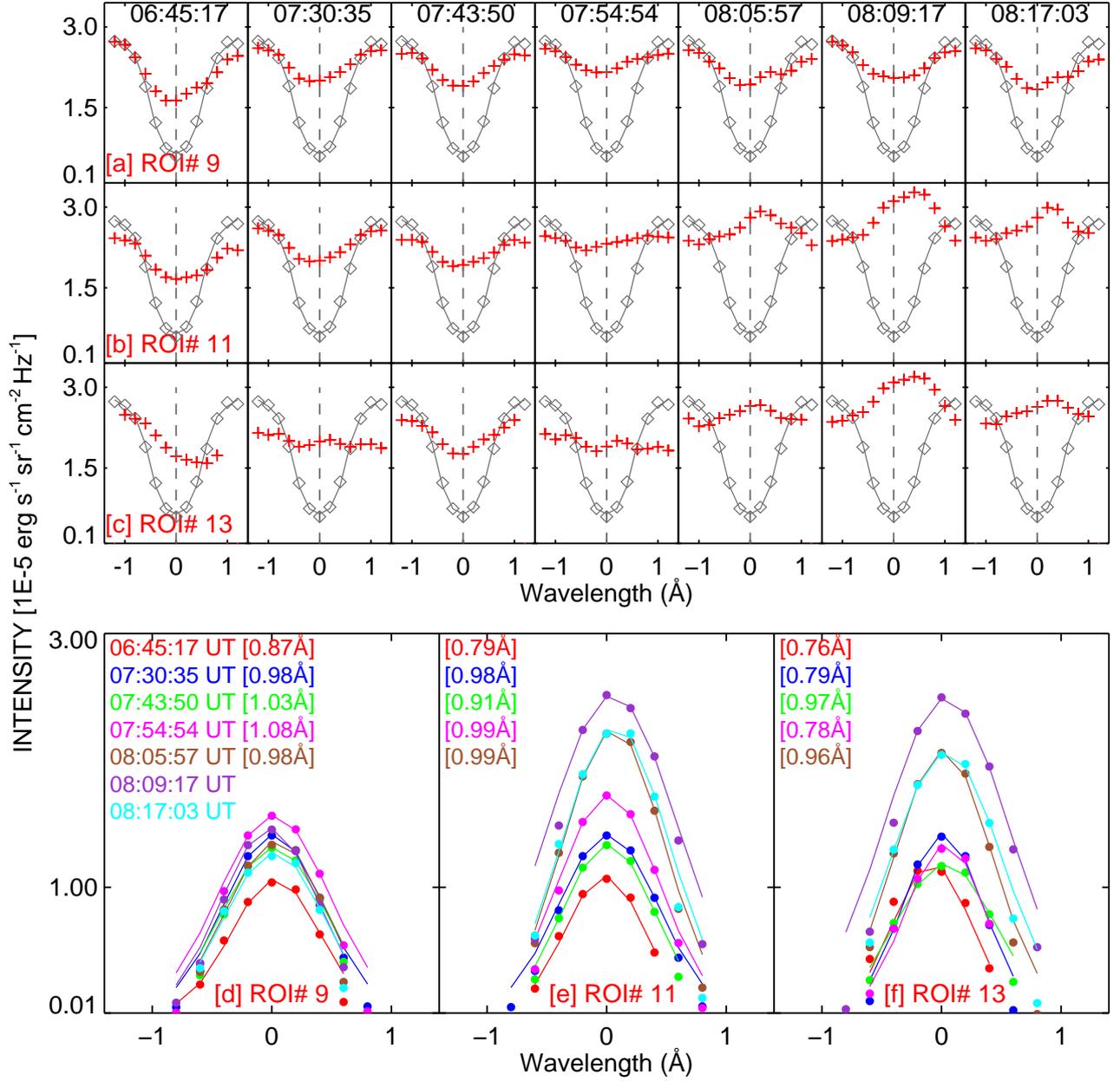

**Figure 6.** [a]-[c]: Time sequence of Hα line profiles corresponding to ROIs 9, 11 and 13. The Hα line profile of the quiet-sun is also shown in grey while vertical dotted line indicates the position of line-center. The net emission profile for the ROIs 9, 11, and 13 is plotted in panels [d]-[f], respectively with coloured dot symbols while the resultant gauss-fit (full line) is also over-plotted with the same color. Different colors represent different time instances of observations and annotated in the panel [d] while the full-width-half-maximum of the respective best-fit gaussian function is also noted for all the ROIs considered in the investigation.

A comparison of the excess emission profile ($I_{excess}$) corresponding to ROI 9 (Figure 7) is made with the SXR emission in 6-12 keV. SXR intensity peak at 07:31 UT can be correlated with the first prominent maxima (P1) among several excursions during the precursor phase in the time profile of Hα line center as well as wings. We note a relative time delay in the peak time of the Hα emission with respect to that in the SXR intensity profile to be approximately

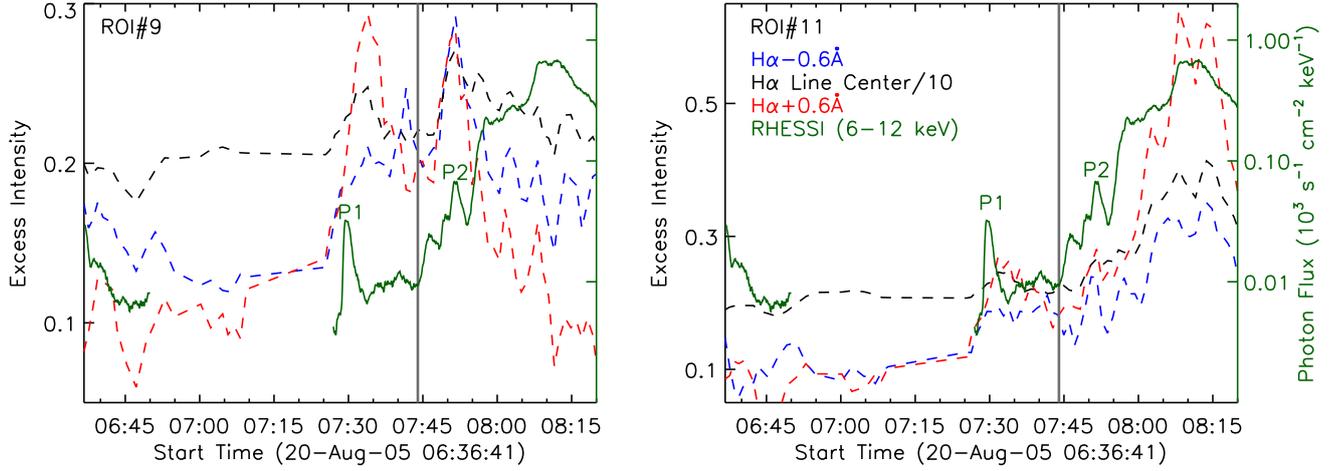

**Figure 7.** Temporal evolution of excess emission in Hα line centre as well as in the wings at ± 0.6 Å for ROIs 9 and 11. I$_{excess}$ corresponding to Hα line center wavelength is divided by 10 in order to match the dynamic range of the excess emission in the wings. X-ray photon flux in 6-12 keV energy band, recorded by *RHESSI* is plotted in green and scaled as per right Y-axis. Two prominent bumps at 07:33 and 07:51 UT are denoted by 'P1' and 'P2' providing the comparative evolution of emission in the Hα line profile with that in the 6-12 keV SXR emission. Vertical line denotes the onset of the main phase of the flare.

180 seconds. The uncertainty in the relative time delay, due to the moderate temporal cadence of images made available from MSDP spectrograph can be as large as ∼ 60 seconds. On the contrary, the second peak (P2) in the Hα time profile, corresponding to the impulsive phase of the flare, appeared without measurable delay compared with the respective SXR peak. Relatively quick response of the Hα emission to the SXR emission during the impulsive phase than that during the precursor phase implies the chromospheric heating through non-thermal electron in this phase (Radziszewski et al. 2011). On the other hand, another slow heating mechanism during the precursor phase may be responsible for delayed response of the Hα emission originating from the chromospheric height. Similarly the investigation of intensity excess in the ROI 11, shown in the right panel of the Figure 7, resulted in a well-correlated temporal evolution of emission in the Hα line center as well as in the wings with the SXR profile. We have not been able to derive any noticeable delay between the maxima of Hα emission profile with that of the SXR time profile during the main phase of the flare.

Additionally, the plots shown in Figure 7 revealed asymmetric responses of the red and blue wings of the profile during various phases of the flare at different locations on the flare ribbon. Until 07:10 UT during the precursor phase, ROI 9 exhibits comparatively stronger emission in the blue wing than that in red wing while the same is not distinctly depicted by the ROI 11. On the contrary, since the onset of the peak P1, both the ROIs exhibit red asymmetry. Red asymmetry in the Hα line profiles derived from the ROI 11 since the onset of main phase of the flare may be caused by the non-thermal electrons, considering the fact that we did not find measurable delay between the emission peaks in the intensity profiles of the Hα line center and the soft X-ray emission during this phase. On the other hand, due to a delayed response of the Hα emission (∼ 180 s) compared with the SXR emission, the red asymmetry during peak P1 may be originated due to the down-flows comparatively slower than the non-thermal electrons of the impulsive phase, possibly a thermal conduction front. Further, while the blue asymmetry during the onset of the flare has been reported in the past (Heinzel et al. (1994) and reference therein), our observations present the signature of blue asymmetry as early as one hour before the onset of the main phase. One of the most plausible interpretation of such as prolonged blue asymmetry in the early stage of the flare may be due to the presence of a moving absorbtion feature such as filament neighbouring the ROIs. It is to note that temporal evolution of the emission in the blue and red-wing corresponding to ROI 9 reveal distinctive nature than that derived from ROI 11 revealing non-uniform chromospheric response corresponding to different locations of the flare ribbon. Therefore, we further investigate the morphology and dynamics of the emission and absorption features. In this regard, we first prepare a time sequence of composite blue-wing images by summing the co-temporal MSDP images recorded in Hα-0.2, Hα-0.4 and Hα-0.6 Å wavelengths. Similarly, the red-wing composite images have been prepared by adding the co-temporal MSDP spectra-images recorded in Hα+0.2, Hα+0.4 and Hα+0.6 Å wavelength positions of the Hα line. Sequence of the blue- and red-wing composite images at 06:36:41 UT (first MSDP record), 07:33:45 UT (corresponding to P1), and at 07:51:35 UT (corresponding to peak P2) is shown in the figure 8.

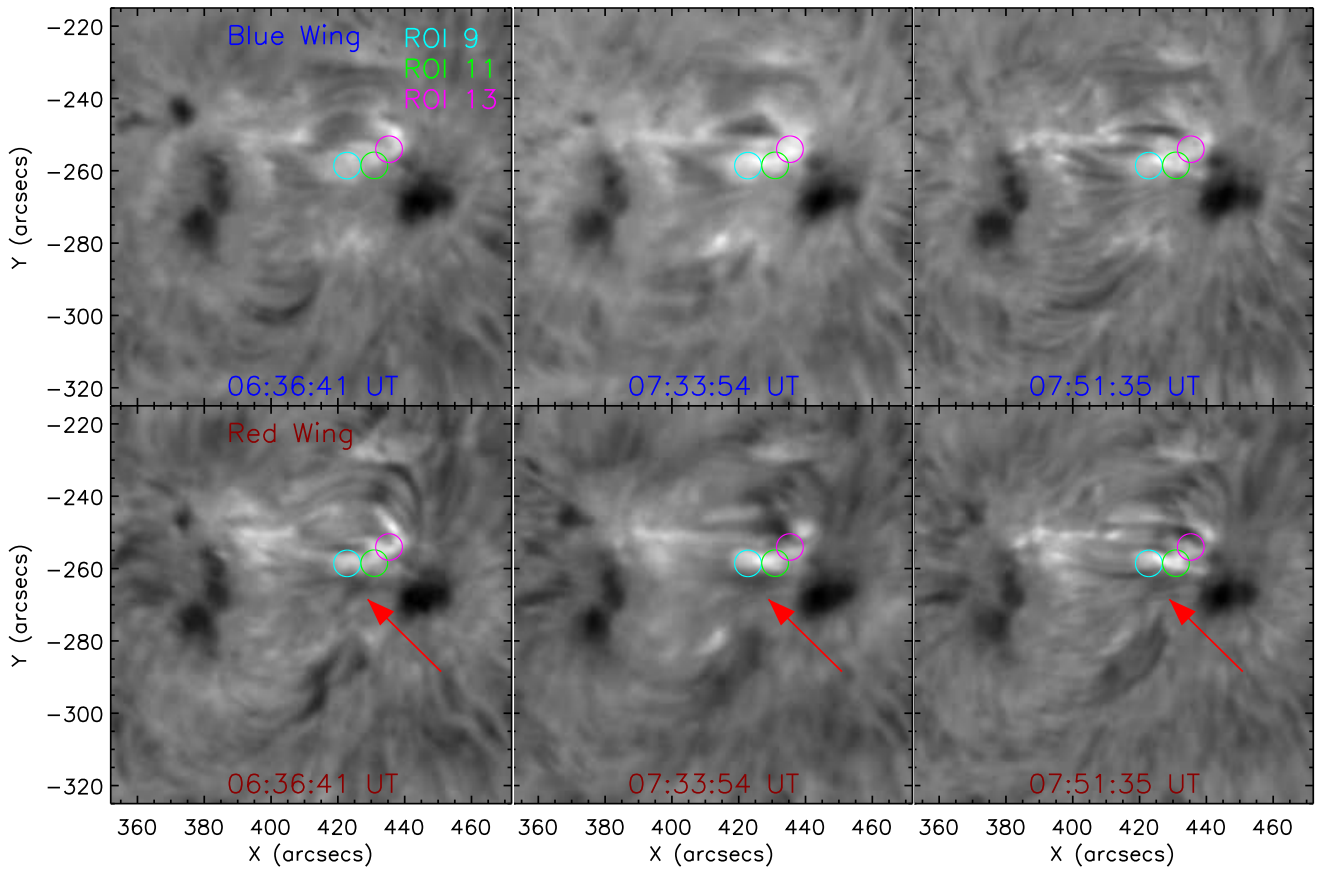

**Figure 8.** Sequence of MSDP composite images representing the blue-wing and the red-wing of the Hα line depicting the morphology and dynamics of the emission and absorption features. Hα composite images at 06:36:41 UT, 07:33:54 UT (corresponding to peak P1 during the precursor phase), and at 07:51:35 UT (peak P2) in the blue-wing (top panel) and red-wing (bottom panel). Red arrows indicate the absorption feature prominently visible only in the red-wing images. Sequence of images corresponding to the blue-wing composite, line-center and red-wing composite of the Hα line profile in the form of a movie is available online.

From the figure 8, we learn that separate portions of a complex filament structure are visible in the blue and red-wing composite images. In particular, in the vicinity of ROI 9, part of a dark filament material (denoted by red arrow) appears only in the red-wing images since the start of the MSDP observations at 06:36 UT. On the contrary, the same structure is not evident in the associated blue-wing images. Therefore, prolonged blue asymmetry during the precursor phase as seen in the Hα line profiles corresponding to ROI 9 may be attributed to a moving absorption feature, possibly referring to draining of the filamentary material similar to that investigated in Huang et al. (2014). Moreover, Graeter & Kucera (1992), in their investigation of Hα line profile during a limb flare found blue-shifted emission profiles during the onset phase of the flare to be originating from the erupting filament. Further, the morphological evolution of the ribbon emission, as shown in figure 8, and the movie, reveals a multi-loop morphology of the active region as well as propagating brightness along the Hα ribbon towards the western (positive) polarity (in agreement to the behaviour of X-ray sources synthesized from *RHESSI* observations (figure 4)). Although the on-disk location of the flaring region concedes the altitude information of flare loops, we propose ROI 9 to be representing the foot-point of a low-lying loop, heated during the precursor phase. Next, propagation of the ribbon brightening implies reconnection to take place sequentially in the overlying loops (Awasthi et al. 2014; Ohyama & Shibata 1997). Foot-point of one such loop, brightened during the main phase of the flare, is represented by ROI 11. This may contribute in the non-uniform nature of line-asymmetry depicted by ROIs 9 and 11.

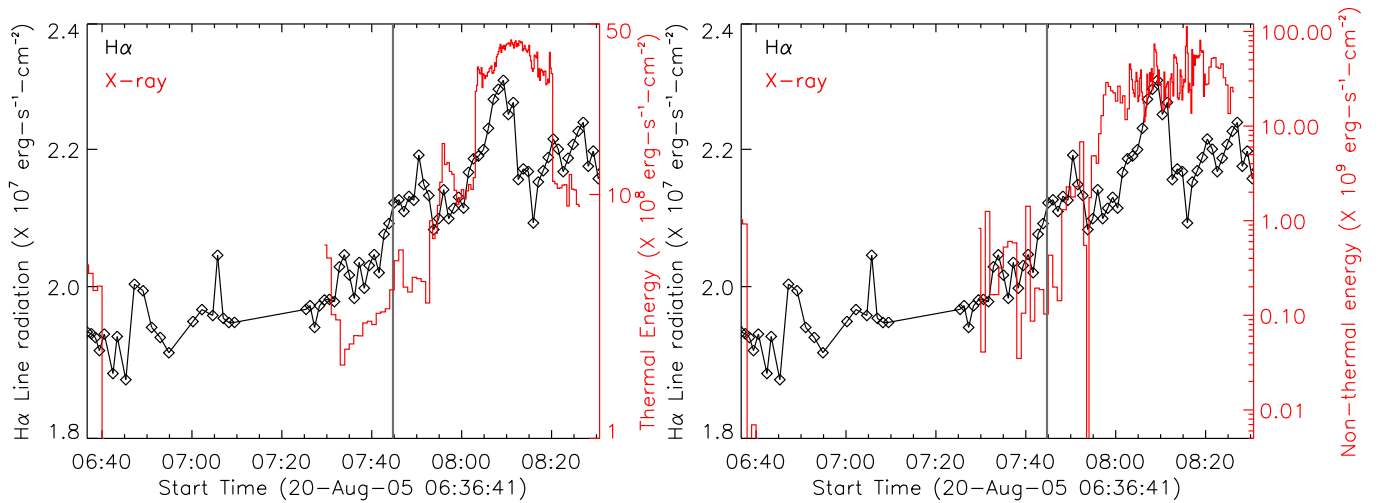

**Figure 9.** Evolution of energy content derived from the Hα line (black) as well as X-ray observations (red). The thermal (left panel) and non-thermal (right panel) energy content estimated using the spectral-fitting of the *RHESSI* observations are plotted and the scaled according to the right Y-axis. Grey line demarcates the onset of the main phase with the precursor phase of the flare.

## 5. ENERGETICS OF THE FLARE

We estimate the energy content in the thermal and non-thermal processes as well as that released from the chromosphere during various phases of the flare. The energy radiated from the flare in Hα line is estimated from the calibrated line profiles (see section 4.1), derived from MSDP observations, during various phases of the flare. For this purpose, we integrate the intensity of the line profile within the wavelength range Hα-0.6Å to Hα+0.6 Å. We intentionally exclude the far wing observations (±0.6 to ±1.2 Å) of the Hα line owing to relatively moderate quality of observations in the said wavelength range. Next, the thermal energetics of the flare is estimated employing the following equation (Saint-Hilaire & Benz 2005).

$$E_{thermal} = 3k_b T \sqrt{EM \cdot V \cdot f} \qquad (2)$$

Here $k_b$ refers to boltzmann constant. T is the temperature of the flaring plasma while V represents the volume of the flaring region, and the same has been approximated to be $A^{3/2}$ where A is the area representing 20% of the maximum intensity in the 195 Å images. 'f' corresponds to the filling factor and considered to be equal to unity in this case. We employ the temperature and emission measure of the plasma, derived by the spectral fit of the *RHESSI* observations to estimate the thermal energetics corresponding to flare observation. In the left panel of Figure 9, the time evolution of the thermal energy content as well as the energy emitted in the Hα waveband is plotted. We have also derived the non-thermal energy content during various phases of the flare with the IDL routine 'calc_nontherm_electron_energy_flux.pro' available in SPEX package of Solarsoft, in which NTE beam parameters estimated from the spectral-fit of *RHESSI* spectra have been supplied as input. Temporal evolution of the non-thermal energy content is shown in the right panel of Figure 9. In Table 1, the maximum values of the energy content derived within various wavebands during the precursor and main phase of the flare are enlisted.

**Table 1.** Maximum energy emitted during the precursor and main phase of the flare

| Energy content | Precursor Phase (erg s$^{-1}$ cm$^{-2}$) | Main Phase (erg s$^{-1}$ cm$^{-2}$) |
|---|---|---|
| Hα | $2.05 \times 10^7$ | $2.3 \times 10^7$ |
| Thermal | $4 \times 10^8$ | $4.3 \times 10^9$ |
| Non-thermal | $2.3 \times 10^9$ | $1.1 \times 10^{11}$ |

The energy radiated in the form of Hα line profile during the precursor phase amounts to 80% of that estimated during the main phase of the flare (Table 1). Moreover, non-thermal energy content is found to be relatively higher than the Hα emission by approximately two orders of magnitude during the precursor phase, while by four orders at the maximum of the main phase of the flare. In agreement to this, Canfield et al. (1991) obtained the ratio of Hα flux to the non-thermal flux to be varying in the range of $10^{-3}$ to $10^{-1}$ for the flares of various intensity classes. Further, comparative analysis revealed that the rate of thermal energy release is approximately 1-2 orders lesser than the non-thermal energy release rate, in agreement with that obtained by Awasthi et al. (2014), Warmuth & Mann (2016b), and Aschwanden et al. (2017).

## 6. ONE DIMENSIONAL HYDRODYNAMIC SIMULATION OF THE THERMAL AND NON-THERMAL EMISSION

We synthesize X-ray spectrum during various phases of the flare by applying a one-dimensional hydrodynamic (1D-HD) numerical model, which can best-fit the observations. Our model makes use of the spectral parameters derived from forward-fitting of the observed X-ray spectra to simulate the evolution of the plasma using a modified Naval Research Laboratory (NRL) Solar Flux Tube Model (Mariska et al. 1982; Mariska & Poland 1985; Falewicz et al. 2009). The spectral properties which serve as an input to the aforementioned model include the non-thermal electron beam (NTE) parameters such as, total electron flux, spectral index, and cut-off energy. Further, the loop geometry as determined from the *RHESSI* observations during the main phase of the flare, while from the EIT images during the precursor phase, is provided as an input to the model. The same along with other input parameters is listed in table 2.

**Table 2.** Input parameters of the model during the precursor and main phase of the flare

| Loop parameter | Precursor Phase | Main Phase |
|---|---|---|
| Half-length (cm) | $3.05 \times 10^9$ | $3.05 \times 10^9$ |
| Flux-tube radius (cm) | $1.2 \times 10^8$ | $1.5 \times 10^8$ |
| Initial Pressure at Base (dyn/cm$^2$) | 6 | 8 |

The implementation of a modified hydrodynamic one-dimensional NRL Solar Flux Tube Model is initiated with the assumption that the flare plasma is confined in a rigid and semi-circular loop. A constant strength of the magnetic field is considered to be 200 G, following the estimations made by Aschwanden (2005). Moreover, a constant cross section of the loop, provided as an input to the code, is estimated in conformity from the soft X-ray images made available by *GOES*/SXI instrument as well as the images in 195 Å wavelength from the *SOHO*/EIT observations. The half-length of the flaring loop, derived from the aforementioned observations, is $3.047 \times 10^9$ cm. In the next step of the model, the plasma inside the loop is set to be heated by the energy flux delivered to the loop solely by NTEs (Falewicz et al. 2011; Falewicz 2014). Subsequently, in order to achieve the best conformity between synthetic (calculated) and observed *GOES* intensity flux in the 1-8 Å range (Falewicz 2014), the low-energy cut-off energies (E$_c$) of the electron spectra, provided as an input to the model, is set to vary for each time step. The steady-state solution of the spatial and spectral distributions of the NTEs are derived for each time step of the model using the Fokker-Planck theory (McTiernan & Petrosian 1990).

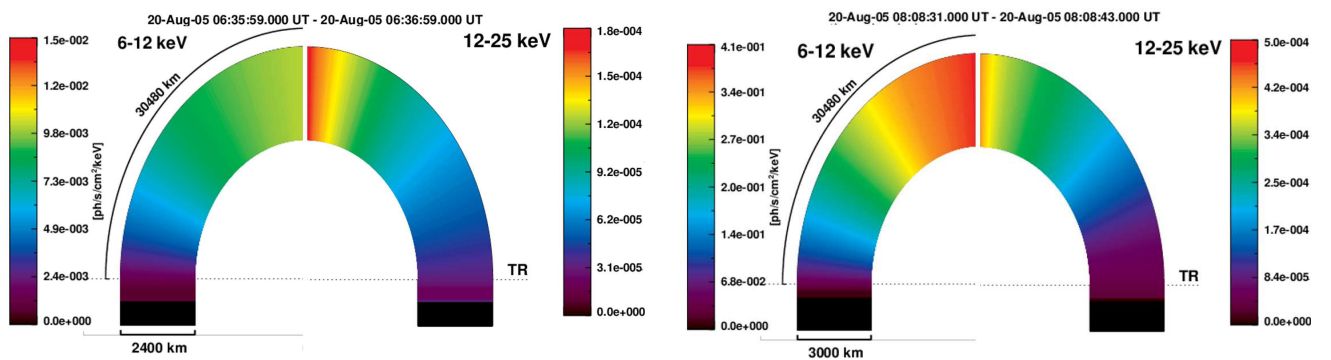

**Figure 10.** Spatial distribution of the emission along the flare loop in 6-12 keV and 12-25 keV energy bands during the precursor and main phases of the flare in the left and right panels, respectively.

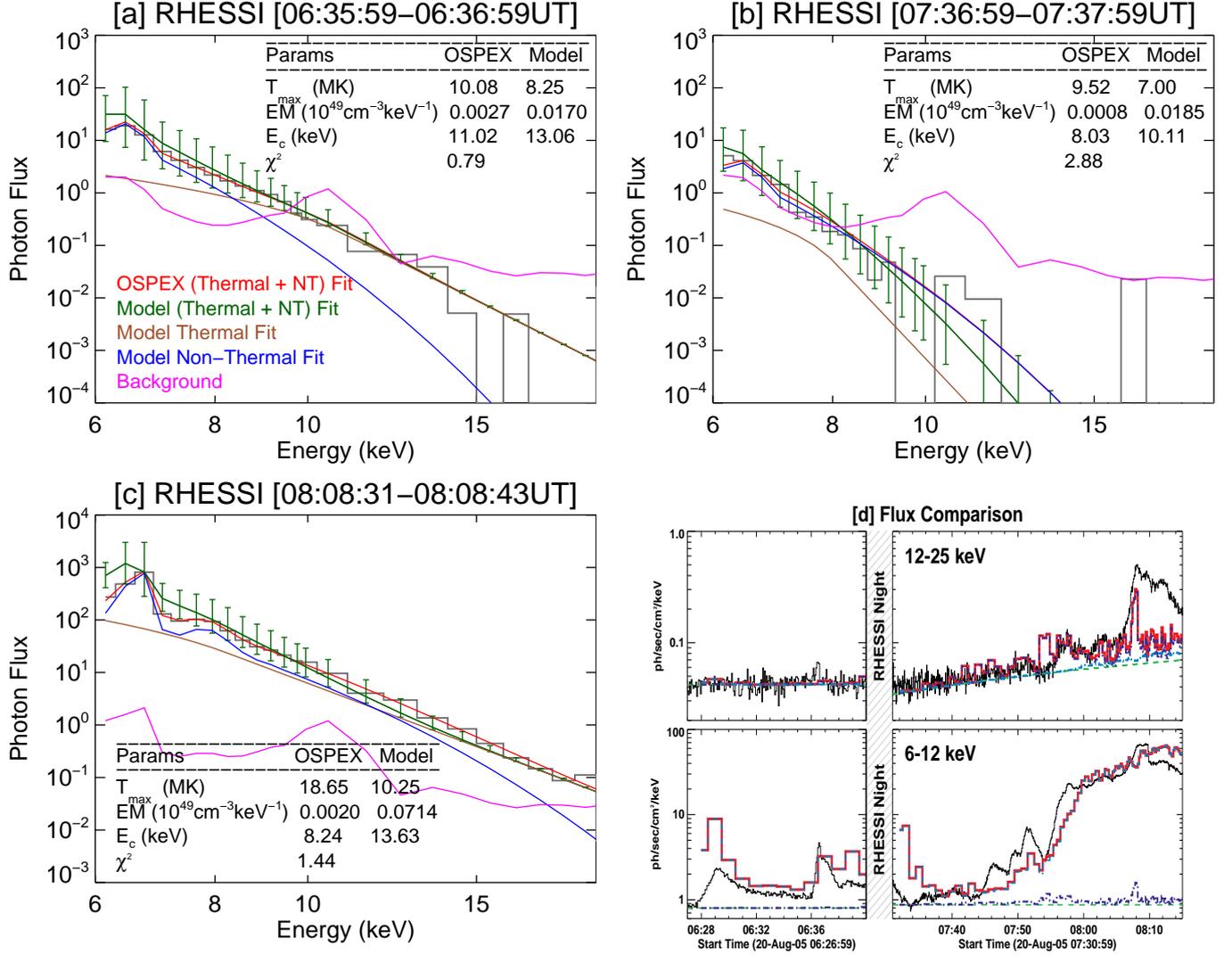

**Figure 11.** Comparison of the spectral evolution of the observed, forward-fitted, and modeled spectra during the precursor phase (panels [a] & [b]), and for the main phase (panel [c]). The modelled thermal spectra are plotted in brown while that corresponding to the non-thermal counterpart are shown in the blue. Observed background from *RHESSI* is shown in magenta color. Uncertainties as resulted by varying the input geometric parameters of the flare loop by 20% of its estimated value are shown in the form of error bars (green) on the respective model X-ray spectrum (green). Panel [d] shows the *RHESSI* observations (black) in 6-12 keV and 12-25 keV (bottom) energy bands, and that estimated (red) by employing model flare plasma parameters. The dotted blue and green lines in this panel represent the calculated and observed backgrounds, respectively.

The spatial distributions of the plasma thermodynamic parameters, as derived from our model, has been applied to derive the X-ray thermal and non-thermal emissions as well as the integral fluxes in the energy ranges of interest. The thermal emission measure of the optically thin plasma was based on the X-ray continuum and line emission calculated using the CHIANTI (version 7.1) atomic code (Dere et al. 1997; Landi et al. 2006). For the plasma temperatures above $10^5$ K, the elemental abundances are based on a coronal abundance set (Feldman & Laming 2000), while below $10^5$ K, photospheric abundances were applied. Ionization equilibrium by Mazzotta et al. (1998) and solar corona abundances for both the lines and continuum have been applied (Feldman et al. 1992). In Figure 10, we show the spatial distribution of emission in 6-12 keV and 12-25 keV energy bands derived from our model at two instance corresponding to the precursor and the main phase, respectively. A comparative investigation of the thermal and non-thermal spectra corresponding to observations as well as that resulted from the HD model is drawn in Figure 11. The X-ray spectrum corresponding to *RHESSI* observations is estimated employing the fit parameters obtained from the application of SPEX package. Employing the aforementioned procedure on the parameters derived from the HD numerical model, we calculate model X-ray spectra. From the X-ray spectrum during the precursor phase, shown in panel [a] of the Figure 11, we find the non-thermal component of the spectra which resulted from the model, to be well below the background level. Similar trend is seen in the panel [b] of the Figure which correspond to the second episode of the precursor emission. Completely developed SXR and HXR emission, significantly above the *RHESSI* background have been derived during the main phase of the flare, as plotted in Figure 11c. The time evolution of the emitted flux in the energy band 6-12 keV and 12-25 keV, resulted from the model, is plotted in the panel [d] of the Figure. The comparison of time evolution of the model flux, drawn in red, with that observed during the precursor phase shows that the flux in 12-25 keV energy band remains below the observed count level. In order to compute the dependence of the model output on input geometric parameters of the flare loop, we ran the HD code for additional cases in which the input geometrical parameters namely loop length (l) and radius (r) have been increased (and decreased) from their original value by 20% constituting four sets ([(l-0.2l, r-0.2r), (l-0.2l, r+0.2r), (l+0.2l, r-0.2r), (l+0.2l, r+0.2r)]). The thermal and non-thermal characteristics of the plasma resulted from the aforesaid four cases of model run is further used to synthesize the model X-ray spectrum. The maximum and minimum deviation of the photon flux of the aforementioned set of X-ray spectra at each energy with respect to the original model X-ray spectra, which is derived employing unaltered set of geometrical parameters (green), is considered as the upper and lower limit of uncertainties, respectively, as shown in the form of error bars (green) in the Figure 11. It may be noted that the uncertainties are not significantly large. Therefore, with the model results, we argue that during the precursor phase, the high-energy photons flux is insufficient to overcome the *RHESSI* detector sensitivity threshold.

## 7. CONCLUSION AND DISCUSSION

We carried out investigation of the chromospheric response and flare energetics during the precursor and main phase of SOL2005-08-20T08:09 event. Spatio-temporal investigation of flare ribbons as seen in the spectra-images, recorded by MSDP in 12 wavelength positions on the H$\alpha$ line profile, revealed a delayed response (180 seconds) of the H$\alpha$ emission compared to the X-ray intensity profile evolution during the precursor phase. On the contrary, no observable delay in H$\alpha$ and X-ray emission peaks could be measured during the main phase of the flare which is suggestive of non-thermal electrons (NTEs) to be the driver of chromospheric heating (Radziszewski et al. 2011) during the impulsive phase. On the contrary, the precursor emission appears to be a consequence of relatively slow heating process. Further, a quantitative investigation of the H$\alpha$ line profiles, estimated from several sub-regions of interest (ROIs) on the flare ribbon, revealed sequential increment in the line-width of the profile during the precursor phase which may be associated with the pressure enhancement in the flare loop (Canfield et al. 1984). In agreement to this, we found increased emission measure (EM) as resulted from the spectral-fit of the observed X-ray emission during the precursor phase.

The investigation of line-asymmetry in the H$\alpha$ emission profiles during the precursor phase revealed comparatively stronger emission in the blue-wing than that in red-wing of H$\alpha$ line corresponding to a few locations on the flare ribbon. While the blue asymmetry during the rise phase of the flare has been reported in the past (Heinzel et al. (1994), Kuridze et al. (2015) and references therein), our observations reveal prolonged blue-wing enhancement as early as an hour before the commencement of the impulsive phase of the flare. In this regard, we investigated morphological and dynamical evolution of the precursor activities in the blue-, and red-wing images of the H$\alpha$ line which revealed the presence of a moving absorption feature (filament material), predominately appeared only in the composite red-wing H$\alpha$ images, in the vicinity and possibly crossing the locations which depict blue asymmetry. Thus, most plausible explanation for the prolonged 'apparent' blue-wing enhancement during the precursor phase is asymmetric absorption (additional to the red-wing) offered by draining filament material (Huang et al. 2014) to the H$\alpha$ line wings. Further, H$\alpha$ line profiles exhibit red asymmetry during the entire flare except during the aforementioned period of blue-asymmetry. During the maximum of the impulsive phase of the flare, the same may be attributed to the non-thermal electrons, owing to the fact that there is no definitive delay between the respective intensity profile peaks. On the other hand, a delayed response of the H$\alpha$ emission compared with the SXR emission during the precursor phase implies that the red-shift in the H$\alpha$ emission during this phase is caused by the down-flows comparatively slower than the non-thermal electrons of the impulsive phase, possibly a thermal conduction front.

Energy content in the H$\alpha$ line during the precursor phase reaches a significant fraction (80%) of that estimated during the main phase. In this regard, we investigate the flare plasma hydrodynamics during the precursor phase from the application of a single-loop one-dimensional model which revealed the presence of a high-energy power-law tail (>10 keV) in the model generated X-ray spectrum, however with the flux values lower than the *RHESSI* background. Therefore, our multi-wavelength diagnostics along with the hydrodynamical modeling of the precursor emission suggests that NTEs flux, although very small, carries sufficient energy to fulfil the requirements of energy budget for plasma heating during this phase. In conclusion, we propose chromospheric response during the precursor phase to be the consequence of two-stage process. In this, reconnection generated NTEs thermalize high in the upper chromosphere during this phase as it contain low energy (Reep et al. 2016). Subsequently, the energy deposited by NTEs is transported down to the lower-chromosphere via conduction. This scenario requires further detailed investigation employing time-correlation analysis of multi-wavelength emission representing various altitudes of the solar atmosphere.

The research leading to these results has received funding from the European Community's Seventh Framework Programme (FP7/2007-2013) under grant agreement no. 606862 (F-CHROMA). A.K.A. acknowledges the support from the Chinese Academy of Science (CAS) as well as the international postdoctoral recruitment program of University of Science and Technology of China. P.R. was supported by the National Science Centre, Poland, under the grant no. UMO-2015/17/B/ST9/02073. A.B. was supported by the grant number 16-18495S of the Czech Funding Agency and by ASI ASCR project RVO:67985815. R.L. acknowledges the support from NSFC 41474151, NSFC 41774150, and the Thousand Young Talents Program of China. This investigation made use of the data freely available from various space-based observatories namely *SDO*, *RHESSI*, *SOHO*, *HINODE*, and *GOES*. The numerical simulations were carried out using resources provided by the Wroclaw Centre for Networking and Supercomputing (http://wcss.pl), grant no. 303. The authors express sincere thanks to the anonymous referee whose elaborated comments enabled to improve the scientific content and the presentation of the paper.

## APPENDIX

### A. TEMPORAL EVOLUTION OF ACTIVE REGION MAGNETIC FLUX

In view of quantifying the signature of magnetic flux emergence or cancellation during various phases of the flare, we analyze one-minute cadence magnetograms made available by MDI instrument aboard *SOHO* mission during the time interval 05:00 -10:00 UT. The signed magnetic flux corresponding to both the polarities have been derived by employing the technique adopted by Savcheva et al. (2014). In this, we first fit the histogram of the magnetic-field values within the active region with a gaussian function. The noise level in the respective magnetogram, assigned by the full-width-half-maximum of the best-fit gaussian function, is estimated to be varying in the range of 50-90 Gauss for the sequence of the magnetograms analyzed in the present investigation. The pixels having magnetic-field above the noise level have been included in deriving the flux. Moreover, the area foreshortening due to projection effect, and the deviation of the LOS magnetic-field ($B_{los}$) with the radial field ($B_r$) have been taken care by the multiplication of $(sec\ \theta)^2$ where $\theta$ is the angular distance from the disk center. The resulted positive and negative flux is plotted in the appendix figure 1. From the evolution, although the high-time (1-min) cadence evolution contains large noise, a 5-min smoothed evolution (shown in red) reveals a slow but steady increase in the positive and negative flux values until 06:18 UT, closely associated in time with the onset of GOES X-ray precursor enhancements (blue dotted line). Followed to this, the flux cancellation is evident, although much clearer in the positive flux evolution.

### B. CALIBRATION OF THE Hα LINE PROFILE USING DAVID

Line profiles shown in Figure 5 are calibrated with the use of DAVID (David 1961) reference spectrum of the Hα line profile for the quiet sun. In this regard, firstly we prepare a time sequence of quiet sun Hα profiles (x,y,$\lambda$,t) from the MSDP spectra-images by averaging over the counts within two rectangles indicative of quiet sun (figure 5; white box). Next, the ratio ($R[\lambda, t]$) of the observed quiet-sun Hα profile and the DAVID (quiet-Sun) reference profile is calculated which serves as a calibration scale for standardizing observed line profiles in the physical units (erg s$^{-1}$ sr$^{-1}$ cm$^{-2}$ Hz$^{-1}$).

### C. TEMPORAL EVOLUTION OF INDIVIDUAL ROIS

Temporal evolution of emission in the Hα line centre wavelength corresponding to all the ROIs (except ROIs 9, 11 and 13), drawn in Figure 5a are plotted in the appendix figure 2. This reveals that while the main phase of the flare is pronounced in almost all of the ROIs simultaneously, only a few exhibit prominent emission during the precursor phase. Therefore, we made a further in-depth investigation of the temporal and spectral evolution of the emission associated with only ROI 9, 11 and 13 as they comprehend the intensity evolution trend of all the ROIs.

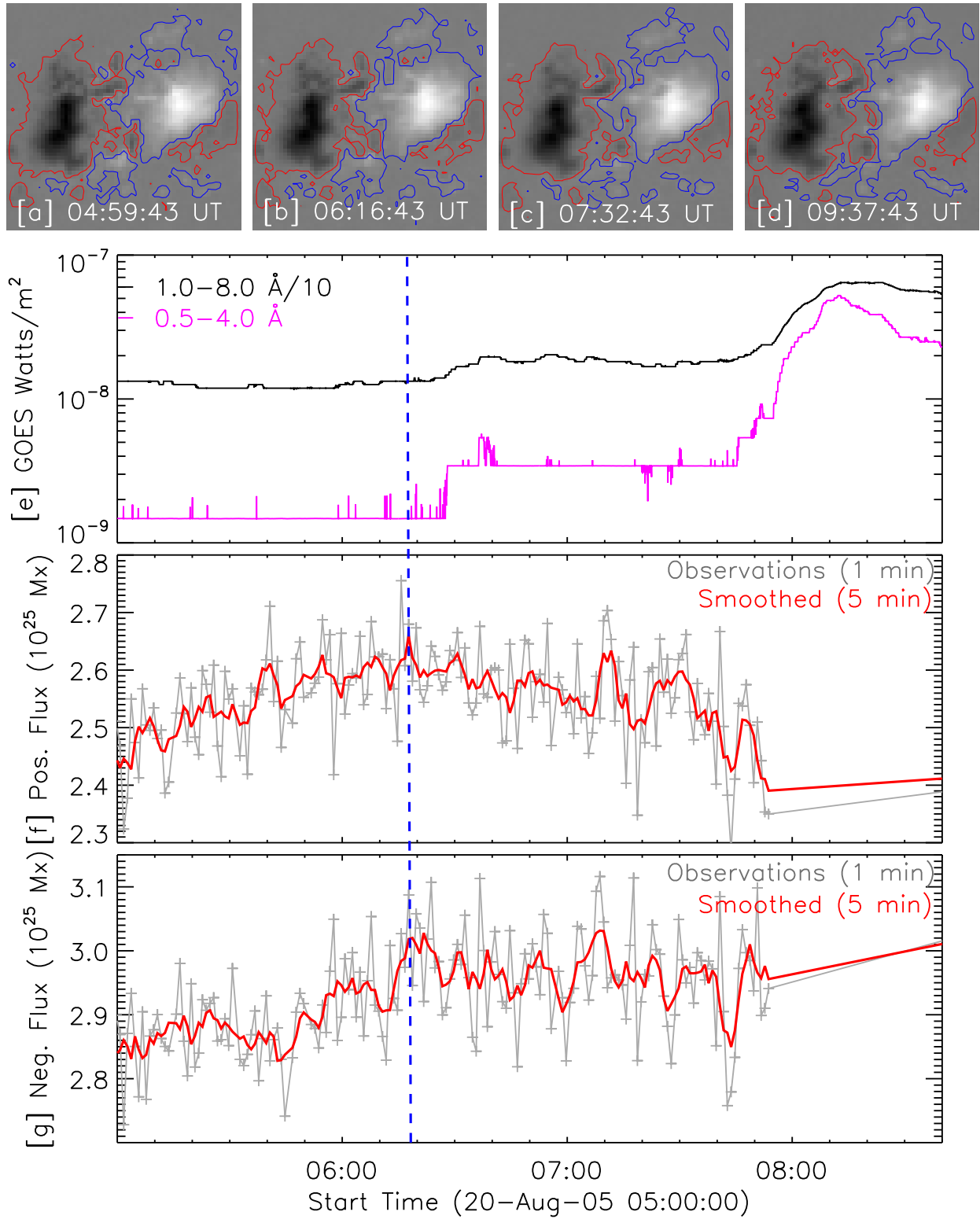

**Appendix Figure 1.** Spatial and temporal volution of Magnetic field parameters. [a]-[d]: Sequence of magnetograms recorded with *SOHO*/MDI at various instances. Temporal evolution of the positive ([f]) and negative ([g]) flux (as derived from the 1-min cadence magnetograms (grey), and 5-min smoothed (red)) is shown along with the X-ray emission ([e]) recorded by *GOES* satellite. Blue dotted line denotes the instance at which positive flux attained its maximum.

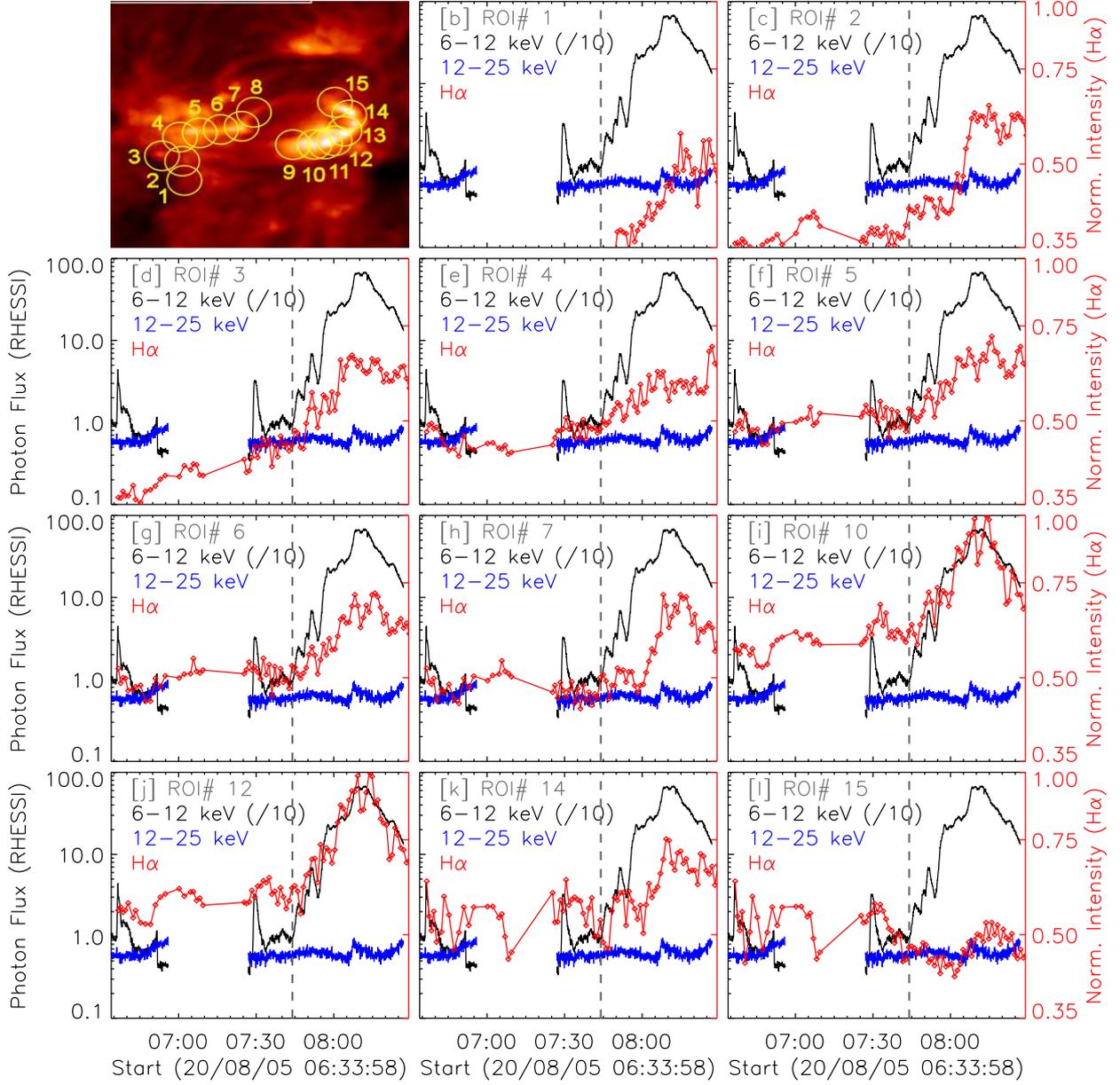

**Appendix Figure 2.** Temporal evolution of emission in the Hα line center wavelength corresponding to the ROIs along the flare ribbon other than the ones which are plotted in the Figure 5.